\documentclass[aps,prb,twocolumn,superscriptaddress]{revtex4-1}

\usepackage {graphicx}

\begin{document}


\title{Thermoelectric properties of Ba-Cu-Si clathrates}


\author{X.\ Yan}
\affiliation{Institute of Solid State Physics, Vienna University of Technology,
Wiedner Hauptstr.\ 8-10, 1040 Vienna, Austria.}
\affiliation{Institute of Physical Chemistry, University of Vienna,
W\"ahringerstr.\ 42, 1090 Vienna, Austria.}

\author {M. X.\ Chen}
\affiliation{Institute of Physical Chemistry, University of Vienna, Sensengasse
8/7, 1090 Vienna, Austria.}
\affiliation{Center for Computational Materials Science, University of Vienna, 
Sensengasse 8/7, 1090 Vienna, Austria}

\author{S.\ Laumann}
\affiliation{Institute of Solid State Physics, Vienna University of Technology,
Wiedner Hauptstr.\ 8-10, 1040 Vienna, Austria.}

\author{E.\ Bauer}
\affiliation{Institute of Solid State Physics, Vienna University of Technology,
Wiedner Hauptstr.\ 8-10, 1040 Vienna, Austria.}

\author{P.\ Rogl}
\affiliation{Institute of Physical Chemistry, University of Vienna,
W\"ahringerstr.\ 42, 1090 Vienna, Austria.}

\author {R.\ Podloucky}
\affiliation{Institute of Physical Chemistry, University of Vienna, Sensengasse
8/7, 1090 Vienna, Austria.}

\author{S.\ Paschen}
\email{Email address:paschen@ifp.tuwien.ac.at}
\affiliation{Institute of Solid State Physics, Vienna University of Technology,
Wiedner Hauptstr.\ 8-10, 1040 Vienna, Austria.}


\date{\today}

\begin{abstract}
Thermoelectric properties of the type-I clathrates Ba$_8$Cu$_x$Si$_{46-x}$
($3.6 \leq x \leq 7$, $x$ = nominal Cu content) are investigated both
experimentally and theoretically. The polycrystalline samples are prepared
either by melting, ball milling and hot pressing or by melt spinning, hand
milling and hot pressing techniques. Temperature-dependent electrical
resistivity, $\rho(T)$, and the Seebeck coefficient, $S(T)$, measurements
reveal metal-like behavior for all samples. For $x = 5$ and $6$, 
density functional theory calculations are performed for deriving the enthalpy of
formation and the electronic structure which is exploited for the calculation
of Seebeck coefficients and conductivity within Boltzmann's transport theory.
For simulating the properties of doped clathrates the rigid band model is
applied. On the basis of the density functional theory results the
experimentally observed compositional dependence of $\rho(T)$ and $S(T)$ of
the whole sample series is analyzed. 
The highest dimensionless thermoelectric
figure of merit $ZT$ of 0.28 is reached for a melt-spun sample at $600^{\circ}$C. The
relatively low $ZT$ values in this system are attributed to the too high
charge carrier concentrations. 
\end{abstract}



\maketitle

\section{Introduction}\label{intro}
Intermetallic clathrates have been investigated extensively in recent years as
promising thermoelectric materials \cite{Sla1995}. They crystallize in several
structure types \cite{2006Rog}, of which the type I clathrates A$_8$E$_{46}$ (A =
alkali, alkaline earth, or rare earth element, E = group IV element) have
hitherto shown the best thermoelectric performance. The unique features of the
clathrate crystal structure are generally regarded as ideal realization of the
phonon glass -- electron crystal (PGEC) concept \cite{Sla1995}. Clathrates are
viewed as Zintl compounds \cite{2007Kau, 2010Tob}: the covalently bonded
framework atoms (E) use the valence electrons of the guest atoms (A) to satisfy
the octet rule and to acquire closed-shell electronic configurations. Simple
electron counting based on the Zintl concept \cite{1939Zin} frequently works
well \cite {1996Kau, 2002Sev, 2007Kau, 2010Tob} for understanding the
thermoelectric properties of clathrates. Previously, most investigations focused
on clathrates with heavy framework atoms (e.g., Ge- or Sn-based clathrates)
which typically are partially substituted by other heavy elements (e.g., Ga, In,
or transition metal elements). Comparably little work has been done on Si-based
clathrates \cite{2008Mel, 2009Nas, 2005Con, 2006Con, 2008Con, 2005Jau, 2011Mug, 2011Tsu}, despite
the technological interest in this low-cost element. Early interest in
Si-based clathrates was triggered by the discovery of superconductivity in
Na$_2$Ba$_6$Si$_{46}$ with a critical temperature $T_c = 4$~K
(Ref.\,\onlinecite{1995Kaw}). Later on, investigations on
Ba$_8$TM$_x$Si$_{46-x}$ (TM: transition metal element, TM = Cu, Ag, Au)
\cite{1998Her, 2000Yam, 2003Ter, 2004Fuk, 2005Yan} demonstrated the
stabilization of the crystal structure by TM substitution, allowing for a simple
synthesis in an arc furnace. However, this led to a suppression of $T_c$
which, from band structure calculations, could be attributed to a change of the
density of states near the Fermi level \cite{2005Kam, 2005Tse, 2005Li}. Interest
in the thermoelectric properties of TM-containing clathrates also arose from
band structure calculations which indicated that Ba$_8$\{Cu,Ag,Au\}$_6$Si$_{40}$
clathrates are degenerate $p$-type semiconductors with large Seebeck
coefficients at room temperature \cite{2004Aka, 2005Aka, 2005Tse}. This
motivated us to investigate the thermoelectric properties of
Ba$_8$Cu$_x$Si$_{46-x}$  in detail.\\
Recently, we investigated the phase equilibrium of clathrates in the Ba--Cu--Si system at 
$800^{\circ}\mathrm{C}$ \ and its crystal structure \cite {2010Yan}. 
On the other hand, first-principles methods have made considerable successes in studies of structural and
electronic properties of materials. 
Moreover, the Boltzmann transport theory has been extensively applied to investigate 
electronic transport properties such as electrical conductivity and Seebeck coefficient.
A method that hybridize the first-principle methods and Boltzmann theory would no doubt benefit 
understanding of thermoelectric properties of materials, 
which has proved to be quite successful in our recent work\cite{zeiringer_phase_2011}.  
In this paper, we present a comprehensive study on the thermoelectric properties of 
Ba$_8$Cu$_x$Si$_{46-x}$ ($3.6 \leq x \leq 7$, $x$ = nominal Cu
content) in combination with density functional theory 
(DFT) calculations and the semi-classic Boltzmann transport equations.

\section{Experimental}\label{exp}
\subsection{Sample preparation and characterization}\label{prep}
A series of clathrates, with the nominal compositions Ba$_8$Cu$_x$Si$_{46-x}$
($x = 3.6$, 3.8, 4.2, 4.4, 4.6, 5, 6, 7, denoted by HP01-HP08), was prepared in
a high-frequency induction furnace from high-purity elements (more than 99.9
mass\%), as described previously \cite{2010Yan}. The samples were subsequently
annealed ($800^\circ$C, 15 days), ball milled and hot pressed ($T = 800^\circ$C,
$P = 56$~MPa, time = 2 hours). A sample with $x = 6$ was chosen for
melt spinning \cite{Paten, 2008Pas} with wheel speeds of 1500 r/min and 2500
r/min (samples MS1500 and MS2500).

X-ray powder diffraction (XPD) data for all samples were collected using a
Siemens D5000 diffractometer (Cu K$_{\alpha 1, 2}$, $10^\circ \leq 2\theta \leq
110^\circ$) and processed with the Rietveld method using the program FULLPROF
\cite{1990Rod}. The crystallite size of hot pressed samples was evaluated from
the X-ray diffraction patterns using the MDI jade 5.0 software (Materials Data
Inc., Liverpool, CA). Annealed ($300^\circ \mathrm {C}$, 1 day) pure Si powder
was used for the evaluation of the peak broadening due to the X-ray instrument.
The bulk density of the hot pressed samples was measured by the Archimedes
method.

The compositions of the hot pressed samples were determined by energy dispersive
X-ray spectroscopy (EDX) in a scanning electron microscope (SEM) operated at 20
kV (Philips XL30 ESEM; JEOL JSM-T330A, probe size: 1$\mu$m). 

\subsection{Physical properties}\label{prop}
The electrical resistivity and the Seebeck coefficient were measured with a
ZEM-3 (ULVAC-Riko, Japan). The thermal conductivity was calculated from the
thermal diffusivity $D_t$, measured by a laser flash method with a
Flashline-3000 (ANTER, USA), the specific heat $C_p$ and the bulk density $D$
using the relation $\kappa  = D_t C_p D$. Hall effect measurements were
performed with a standard ac technique in a physical property measurement system
(PPMS, Quantum Design) in the temperature range from 2 to 300 K in magnetic
fields up to 9 T.

\subsection{Computational details}\label{comp}
Density functional theory (DFT) calculations were carried out by the Vienna \it ab initio
\normalfont Simulation Package (VASP) \cite{1996Kre,1996Kre2}.
The exchange-correlation functional 
is approximated within the generalised gradient approximation
using the parametrization of Perdew, Burke, and Ernzerhof (PBE). 
The electron-ion interaction is treated
within the framework of Bl\"{o}chls projector augmented wave method
\cite{paw1,paw2}.  The valence state configurations for the construction
of the pseudopotentials included  $5s^{2}5p^{6}6s^2$ states for Ba,
$3p^{6}3d^{10}3s^{1}$ states for Cu, and $3s^{2}3p^{2}$ states for Si. 
For all calculations a $5\times5\times5$
$\mathbf{k}$ point grid according to Monkhorst and Pack \cite{PhysRevB.13.5188}
was used to sample the Brillouin zone.

Transport properties for the electrons were modeled
within  the semi-classical Boltzmann transport theory in which
the Seebeck coefficient tensor ($S$) and electrical resistivity ($\rho$) are defined as
 \begin{equation}
\rho=\sigma^{-1} \hspace{4mm} \sigma(T,E_F) = e^2K_0 
  \label{conductivity}
  \end{equation}
 \begin{equation}
S(T,E_F) =  \frac{K_1}{eTK_0}
  \label{seebeck}
  \end{equation}
Here $K_n$ is related to electronic structures by
\begin{eqnarray} \nonumber
K_n &=& \frac{1}{4\pi^3}\frac{\tau}{\hbar}\sum_{i,\mathbf{k}}\mathbf{v(i,\mathbf{k})}
\mathbf{v(i,\mathbf{k})}(E(i,\mathbf{k})-E_F)^n  \\
 &&  (-\frac{\partial f_{E_F}(T,E(i,\mathbf{k}))}{\partial E}) .
  \label{k_n}
\end{eqnarray}
The vectors $\mathbf{v}=\partial E(i,\mathbf{k})/\partial \mathbf{k}$
represent the velocities of the electrons as defined by the derivative of
the band energy.  $\tau$, $f$ and $E_F$ are the relaxation
time, the Fermi function, and the Fermi energy, respectively.  Within
Boltzmann¿s transport theory effects of electron-electron and electron-phonon
scattering are merged  into the relaxation time $\tau$.  Because a
first-principles calculation  of $\tau$ is not feasible  for systems with
large numbers of atoms per unit cell, $\tau$ was considered  as an empirical
parameter  by fitting  to one selected  experimental value at a given
temperature, as described below.

It should be noted that the cubic symmetry of Ba$_8$Si$_{40}$ is maintained for
Ba$_8$Cu$_6$Si$_{40}$ but broken when modeling Ba$_8$Cu$_5$Si$_{41}$. 
Therefore, the Seebeck tensor has now
unequal components. Because the experimental samples were produced at higher
temperatures the measured structures are always cubic with some random site
occupations. In order to enable the comparison with experiment the symmetry
average of the calculated Seebeck and conductivity tensor is made, resulting
in only one component for each physical property.
Concerning the transport properties of Ba$_8$Cu$_5$Si$_{41}$ two sets of
calculations have been made, namely for a geometrically fully relaxed structure and for a
structure denoted as  frozen-Ba$_8$Cu$_5$Si$_{41}$,
for which the high-symmetry structure of Ba$_8$Cu$_6$Si$_{40}$ was chosen and no
relaxation was allowed.
%
The actual calculations of the transport properties were made by an adapted version of
the package BoltzTrap \cite{madsen_boltztrap._2006}. 
For this purpose the Kohn-Sham energy eigenvalues were
generated on a very dense  $25\times25\times25$ $\mathbf{k}$-point grid.
\section{Results and discussion}
\subsection{Sample purity in the hot pressed samples and structural details of
clathrates}\label{structure}
The phase constitutions of all hot pressed samples agree well with the
isothermal section of the system Ba--Cu--Si at $800^\circ \mathrm {C}$ (the
homogeneity range of the clathrate at $800^\circ \mathrm {C}$: $3.4(1)\leq
x_{Cu} \leq 4.8(1)$, Ref.\,\onlinecite{2010Yan}), except for a very small amount
of Si (and/or of holes) that is uniformly dispersed in the main clathrate phase
(even for the samples with $x \geq 5$). The compositions from EDX measurements
are close to the nominal compositions for the samples HP01-HP05 which correspond
to the single phase range of clathrate in the phase diagram. For the samples
HP06-HP08, a composition of Ba$_{8.1}$Cu$_{4.9}$Si$_{41.0}$ is determined.
Table~\ref{Tab:tab1} summarizes the nominal Cu content, the Cu contents
determined from EDX and from the XRD refinements, as well as the crystallite
sizes and densities of all specimens.

\begin{table}
 \caption{\label{Tab:tab1} Hot pressed Ba$_8$Cu$_x$Si$_{46-x}$ samples with
their nominal Cu content ($x_{nom}$), the Cu content determined from EDX
measurements
($x_{EDX}$) and from the XRD refinements ($x_{ref}$), the bulk density ($D_b$) and
the crystallite size ($S_c$).}
 \begin{ruledtabular}
 \begin{tabular}{c c c c c c}
  Name & $x_{nom}$ & $x_{EDX}$ & $x_{ref}$ & $D_b$ (\%) & $S_c$ (nm) \\
\hline
HP01 & $3.6$ & 3.5 & 3.61 & $92.7$ & $165(15)$\\
HP02 & $3.8$ & 3.7 & 3.85 & $92.8$ & $125(10)$\\
HP03 & $4.2$ & 4.2 & 4.29 & $92.0$ & $120(5)$\\
HP04 & $4.4$ & 4.3 & 4.38 & $88.7$ & $130(10)$\\
HP05 & $4.6$ & 4.5 & 4.55 & $91.8$ & $135(15)$\\
HP06 & $5.0$ & 4.9 & 4.72 & $95.0$ & $140(15)$\\
HP07 & $6.0$ & 4.9 & 4.88 & $96.6$ & $155(5)$\\
HP08 & $7.0$ & 4.9 & 4.98 & $92.5$ & $160(10)$\\
MS1500 & $6.0$ & 4.8 & 4.84 & 90.4 &  \\
MS2500 & $6.0$ & 4.8 & 4.82 & 88.1 &  \\
 \end{tabular}
 \end{ruledtabular}
\end{table}

\begin{figure}
 \includegraphics[width=0.4\textwidth, bb=0 0 500 403]{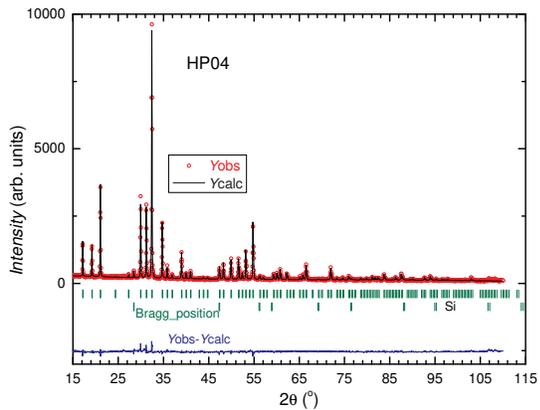}%
 \caption{\label{Fig:fig1} (Color online) The Rietveld refinement of the powder
X-ray diffraction data from HP04 shows the main clathrate phase with a small
amount of Si. The microstructure of this sample (insert) shows the main grey
phase (clathrate phase) with black spots (Si and/or holes).}
\end{figure}

Rietveld refinements of the X-ray diffraction data were performed with an
initial model derived from the single crystal X-ray diffraction data of
Ba$_8$Cu$_5$Si$_{41}$ (Ref.\,\onlinecite{2010Yan}). Here, Ba atoms fully occupy
the $2a$ and $6c$ sites in the crystal structure of the type-I clathrate, Cu
atoms share the $6d$ site with Si, and the remaining sites are occupied by Si
atoms. For this model the Rietveld refinements of the data for all samples have
high reliabilities and yield reasonable structure parameters (see
Table~\ref{Tab:tab2}). Unlike for the case of several Ge-based TM clathrates
\cite{2008Mel, 2009Mel,2010Ngu,2007Mel,2007Mel2,2007Mel3,2009Pro, 2010Koz, 2011Zha}, here no
vacancy could be identified at the $6d$ site within experimental resolution.
Figure~\ref{Fig:fig1} shows the Rietveld refinement for the sample HP04 as an
example. 

\begin{table*}
 \caption{\label{Tab:tab2} Structure data for Ba$_8$Cu$_x$Si$_{46-x}$ ($3.6 \leq x \leq 7$). The data have been standardized by the program Structure Tidy \cite{1994Par}.}
 \begin{ruledtabular}
 \begin{tabular}{l c c c c}
 \bf Name & \bf HP01  & \bf HP02 & \bf HP03 & \bf HP04  \\
\hline
$x_{ref}$ & 3.61 & 3.85 & 4.29 & 4.38 \\
$a$ (nm), Guinier & 1.03272(3) & 1.03273(5) & 1.03278(3) & 1.03277(6) \\
$R_F=\sum|F_o-F_c|/\sum F_o$ & 0.105  &  0.122  &  0.082  &  0.075 \\
$R_I=\sum|I_o-I_c|/\sum I_o$ & 0.085  &  0.081  &  0.073  &  0.060 \\
Ba1, in $2a$ ($0,0,0$) $B_{eq}$ ($B_{iso}$) 10$^2$(nm$^2$) & 0.41(9) & 0.62(9) & 0.47(9) & 0.41(7) \\
Ba2, in $6c$ ($\frac{1}{4},0,\frac{1}{2}$) $B_{eq}$ ($B_{iso}$) 10$^2$(nm$^2$) & 1.71(9) & 1.72(9) & 1.63(8) & 1.57(5) \\
M1, in $6d$ ($\frac{1}{4},\frac{1}{2},0$), Occ.  & 3.61(7)Cu+2.39Si & 3.85(6)Cu+2.15Si & 4.29(6)Cu+1.71Si & 4.38(5)Cu+1.62Si \\
$B_{eq}$ ($B_{iso}$) 10$^2$(nm$^2$) & 0.32(8) & 0.43(9) & 0.42(9) & 0.53(9) \\ 
Si1 in $16i$ ($x,x,x$), $x$ & 0.1855(3) & 0.1854(3) & 0.1852(3) & 0.1851(2) \\
$B_{eq}$ ($B_{iso}$) 10$^2$(nm$^2$) & 0.54(7) & 0.85(9) & 0.59(9) & 0.66(9) \\ 
Si2 in $24k$ ($0,y,z$), $y,z$ & 0.1198(5), 0.3077(5) & 0.1188(4), 0.3091(5) & 0.1189(4), 0.3098(5) & 0.1192(3), 0.3094(3) \\
$B_{eq}$ ($B_{iso}$) 10$^2$(nm$^2$)  & 0.43(9) & 0.25(7) & 0.43(9) & 0.27(9) \\ 
 \end{tabular}
 \end{ruledtabular}
\end{table*}

\begin{table*} [split] 
 \caption{continued}
 \begin{ruledtabular}
 \begin{tabular}{l c c c c}
 \bf Name & \bf HP05 & \bf HP06 & \bf HP07 & \bf HP08  \\
\hline
$x_{ref}$ & 4.55 & 4.72 & 4.88 & 4.98 \\
$a$ (nm), Guinier & 1.03280(3) & 1.03286(4) & 1.03285(5) & 1.03287(4) \\
$R_F=\sum|F_o-F_c|/\sum F_o$ & 0.107  &  0.088  &  0.083  &  0.088 \\
$R_I=\sum|I_o-I_c|/\sum I_o$ & 0.086  &  0.071  &  0.071  &  0.065 \\
Ba1, in $2a$ ($0,0,0$) $B_{eq}$ ($B_{iso}$) 10$^2$(nm$^2$) & 0.49(7) & 0.41(9) & 0.38(9) & 0.40(9) \\
Ba2, in $6c$ ($\frac{1}{4},0,\frac{1}{2}$) $B_{eq}$ ($B_{iso}$) 10$^2$(nm$^2$) & 1.63(6) & 1.55(9) & 1.52(7) & 1.63(8) \\
M1, in $6d$ ($\frac{1}{4},\frac{1}{2},0$), Occ.  & 4.55(5)Cu+1.45Si & 4.72(7)Cu+1.28Si & 4.88(6)Cu+1.12Si & 4.98(5)Cu+1.02Si \\
$B_{eq}$ ($B_{iso}$) 10$^2$(nm$^2$) & 0.65(9) & 0.68(8) & 0.43(9) & 0.86(9) \\ 
Si1 in $16i$ ($x,x,x$), $x$ & 0.1842(2) & 0.1849(3) & 0.1848(3) & 0.1849(3) \\
$B_{eq}$ ($B_{iso}$) 10$^2$(nm$^2$) & 0.68(9) & 0.62(9) & 0.65(9) & 0.68(9) \\ 
Si2 in $24k$ ($0,y,z$), $y,z$ & 0.1203(3), 0.3096(2) & 0.1190(5), 0.3093(5) & 0.1200(4), 0.3095(4) & 0.1195(3), 0.3092(3) \\
$B_{eq}$ ($B_{iso}$) 10$^2$(nm$^2$)  & 0.40(9) & 0.52(9) & 0.28(7) & 0.39(9) \\ 
 \end{tabular}
 \end{ruledtabular}
\end{table*}

\subsection{Thermoelectric properties}\label{TEprop}
\subsubsection{Hot pressed Ba$_8$Cu$_x$Si$_{46-x}$ samples}
\begin{figure}
 \includegraphics[width=0.4\textwidth]{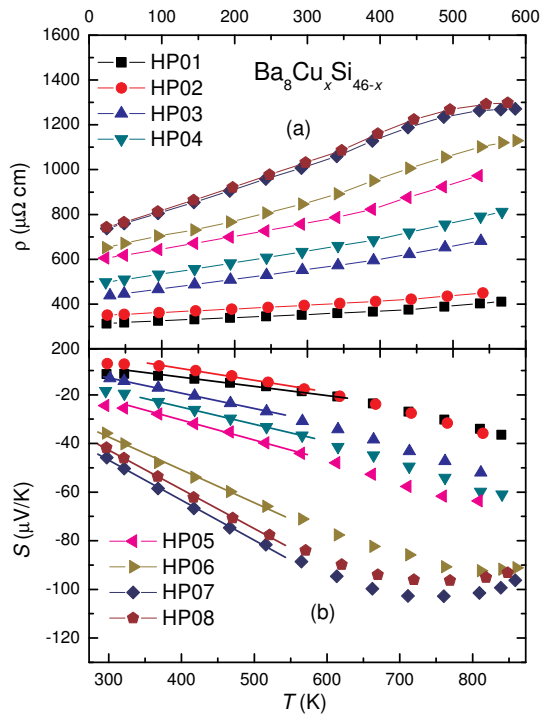}%
 \caption{\label{Fig:fig2} (Color online) Temperature dependent electrical
resistivity, $\rho(T)$, and Seebeck coefficient, $S(T)$, for
Ba$_8$Cu$_x$Si$_{46-x}$ ($3.6 \leq x \leq 7$) plotted as a function of
temperature. The straight lines in $S(T)$ represent linear fits through the
origin.}
\end{figure}

The temperature dependent electrical resistivity, $\rho(T)$, and Seebeck
coefficient, $S(T)$, of the hot pressed samples show a systematic evolution with
the Cu content (Fig.\,\ref{Fig:fig2}). All samples show metal-like behavior.
The negative $S(T)$ values indicate that electrons dominate the transport
properties. In the Cu-rich samples (HP06--HP08), extrema are observed in
$\rho(T)$ and $S(T)$. 

The overall variation of $\rho(T)$ and $S(T)$ with Cu content can be
qualitatively understood within a phenomenological picture of the electronic
band structure of TM containing clathrates \cite{2005Tse}. In all samples, the
Fermi level lies within the conduction band. This explains the metal-like
behavior of $\rho(T)$ and the negative $S(T)$ values. With increasing $x$, the
Fermi level moves towards the edge of the conduction band. Thereby, the charge
carrier concentration $n$ decreases and, consequently, the absolute values of
$\rho(T)$ and $S(T)$ increase. 

According to band structure calculations for a fully ordered structure,
Ba$_8$Cu$_6$Si$_{40}$ should be a $p$-type semiconductor (see Sect.\,\ref{theo}
and Refs.\,\onlinecite{2004Aka, 2005Aka}). The fact that only $n$-type behavior
is found in our sample series is attributed to the formation of competing phases
when $x_{nom}$ exceeds $4.9$. This inhibits a further energetical lowering of
the Fermi level down to the valence band. The energy gap is predicted to
decrease with increasing Cu content (see Sect.\,\ref{theo} and
Refs.\,\onlinecite{2004Aka, 2005Aka}). This is in agreement with the observed
maxima in $\rho(T)$ and $S(T)$ for the Cu-rich samples HP06-08, which signal
the onset of intrinsic conduction above about $500^{\circ}$C.

To a first approximation, for temperatures above the Debey temperature
$\theta_D$ ($\theta_D \approx 150^\circ$C for Ba$_8$Cu$_5$Si$_{41}$,
Ref.\,\onlinecite{2011Yan}), and ignoring the phonon-drag contribution to the
total Seebeck coefficient \cite{1964Cut}, $S(T)$ should vary linearly with
temperature as \cite{1964Bla} 
\begin{equation}
S = \frac {2\pi ^2 k_B^2 m_e}{e \hbar^2(3n\pi ^2)^{2/3}}T
\label{eq:eq1}
\end{equation}
where $n$ is the charge carrier concentration, $m_e$ is the free-electron mass,
$e$ is its charge and the other symbols have their usual meaning. Thus, one
should be able to estimate $n$ from the slopes of linear fits to $S(T)$. These
fits are shown as straight lines in Fig.\,\ref{Fig:fig2}. For high Cu
contents, these fits describe the data quite well in relatively broad
temperature ranges. For low Cu contents, the agreement is poorer. As will be
shown below, this can be attributed to a temperature dependent charge carrier
concentration.

\begin{figure}
 \includegraphics[width=0.4\textwidth]{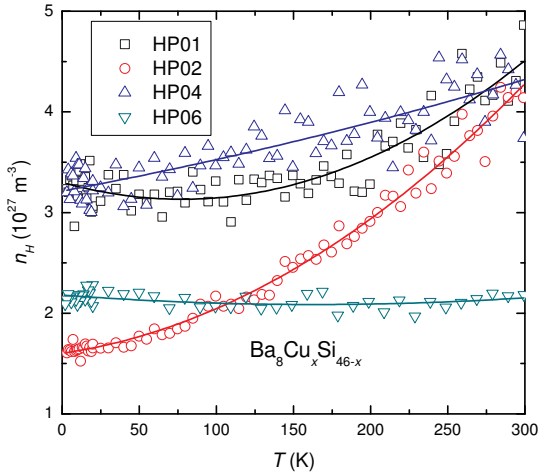}%
 \caption{\label{Fig:fig3} (Color online) Temperature dependent charge carrier
concentration $n_H(T)$ of HP01, HP02, HP04 and HP06 derived from Hall effect
measurements in a one-band model. The lines are guides to the eye.}
\end{figure}

For the samples HP01, HP02, HP04 and HP06 we have also determined the charge
carrier concentrations more directly, by measurements of the Hall coefficient
$R_H(T)$ below 300~K. The charge carrier concentrations calculated using a
simple one-band model $n_H = 1/(eR_H)$, are plotted against temperature in
Fig.\,\ref{Fig:fig3}. For HP06, $n_H(T)$ is almost temperature independent. For
the Cu-poor samples, on the other hand, $n_H(T)$ shows a sizable temperature
dependence.

\begin{figure}
 \includegraphics[width=0.4\textwidth]{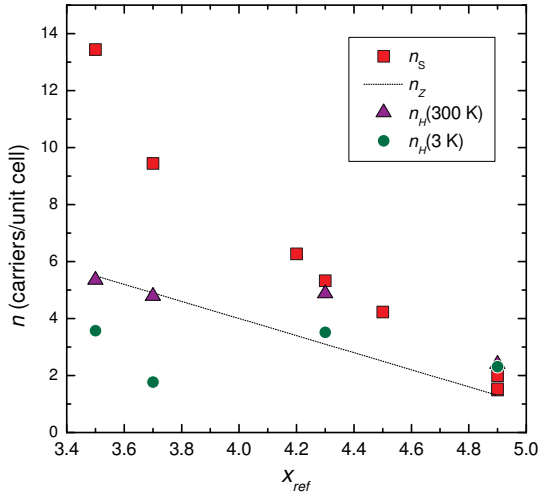}%
 \caption{\label{Fig:fig4} (Color online) Charge carrier concentrations derived
from different methods (see text) vs $x_{EDX}$.}
 \end{figure}

Finally, we can derive the charge carrier concentration expected within a
simple  Zintl electron counting scheme. For Ba$_8$Cu$_x$Si$_{41-x}$, the charge
carrier concentration per unit cell, $n_Z$, varies with $x$ as $n_Z(x) = 16 -
3x$ , if the oxidation states +2, -3, and 0 are assumed for Ba, Cu and
Si\cite{2005Li}, respectively. A comparison of the charge carrier concentration
determined via all above methods is shown in Fig.\,\ref{Fig:fig4}.

\begin{figure}
 \includegraphics[width=0.4\textwidth]{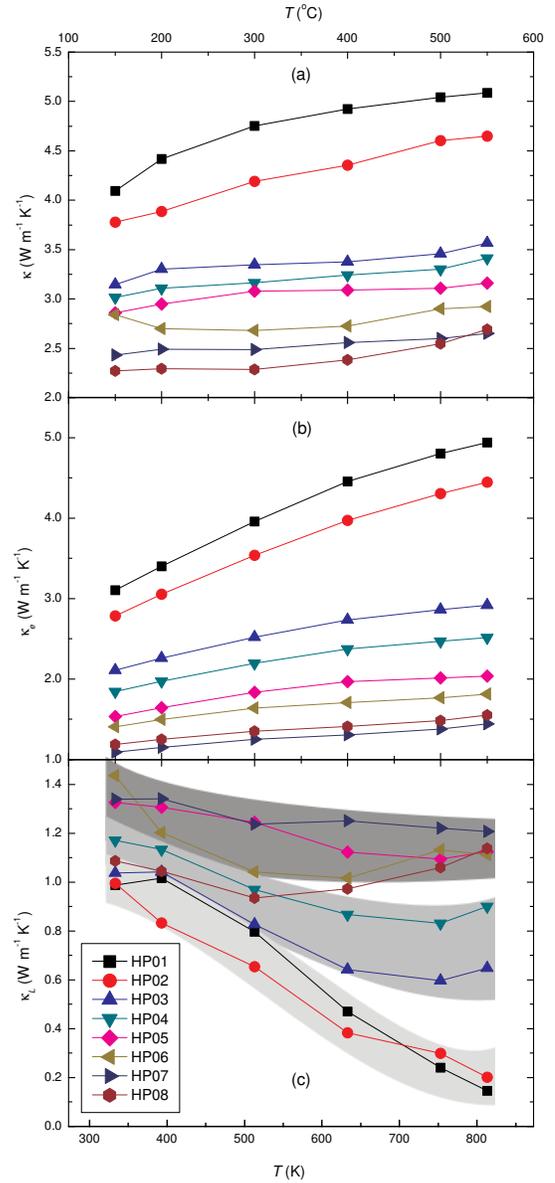}%
 \caption{\label{Fig:fig5} (Color online) Temperature dependent total thermal
conductivity $\kappa(T)$ in (a), electronic contribution $\kappa_e(T)$ in (b)
and lattice contribution $\kappa_L(T)$ in (c) for Ba$_8$Cu$_x$Si$_{46-x}$ ($3.6
\leq x \leq 7$). The gray shaddows in (c) indicated different characteristics in
$\kappa_L(T)$ (see text).}
\end{figure}

The temperature dependent thermal conductivity, $\kappa(T)$, varies
systematically with $x$, as shown in Fig.\,\ref{Fig:fig5}(a). In
Fig.\,\ref{Fig:fig5}(b) and (c), we plot the temperature dependent electronic
component $\kappa_e(T)$ and the lattice component $\kappa_L(T)$, respectively.
$\kappa_e$ is estimated via the Wiedemann-Franz law $\kappa_e = L_0T/\rho$,
where $L_0 = 2.44\times 10^{-8}$~(V/K)$^2$, from the electrical resistivity
$\rho$. $\kappa_L$ is the difference between $\kappa$ and $\kappa_e$. The
dominating contribution to $\kappa$ changes from $\kappa_e$ to $\kappa_L$ with
increasing Cu content, with almost half to half contributions for the samples
HP05 to HP08. For $\kappa_L(T)$ three different characteristics can be
distinguished beyon the experimental uncertainties, as indicated by the gray
shaddows shown in Fig.\,\ref{Fig:fig5}(c).

The Cu-poor samples have the strongest temperature dependence and the lowest
$\kappa_L$ values. With increasing $x_{EDX}$ and thus decreasing charge carrier
concentration (at elevated temperatures), this dependence becomes more gradual
and $\kappa_L$ rises. This trend cannot result from point scattering or boundary
scattering since the crystallite size and bulk density do not vary
systematically with $x_{EDX}$ (Table \ref{Tab:tab1}). Instead suggest that,
besides phonon-phonon scattering that usually is the dominating scattering at
high temperature \cite{1929Pei}, here also electron-phonon scattering
contributes sizably. An alternative explanation for the experimentally observed
trend could be that the presence of more atoms Cu disturbs the resonant
scattering of the guest atoms in the large cage.

\begin{figure}
 \includegraphics[width=0.4\textwidth]{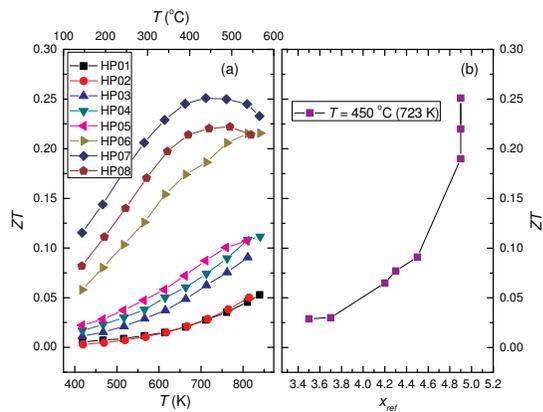}%
 \caption{\label{Fig:fig6} (Color online) (a) Figure of merit $ZT$ for
Ba$_8$Cu$_x$Si$_{46-x}$ ($3.6 \leq x \leq 7$) versus temperature $T$. (b) $ZT$
vs.\ $x_{EDX}$ at $T = 450^\circ$C.}
\end{figure}

The dimensionless thermoelectric figure of merit $ZT = S^2T/\rho\kappa$ is
plotted against temperature in Fig.\,\ref{Fig:fig6}(a). $ZT$ depends sensitively
on $x_{EDX}$, especially for the Cu-rich side (Fig.\,\ref{Fig:fig6}).
Surprisingly, the samples HP06--HP08 with highest $x_{EDX}$ and a small amount
of foreign phases show the highest $ZT$ values in the investigated temperature
range. The highest $ZT$ of 0.25 is reached for sample HP07 at about
$450^\circ$C. Between 350 to $550^\circ$C, $ZT$ remains above 0.23.

\subsubsection{Melt-spun $Ba_8Cu_6Si_{40}$}
\begin{figure}
 \includegraphics[width=0.4\textwidth]{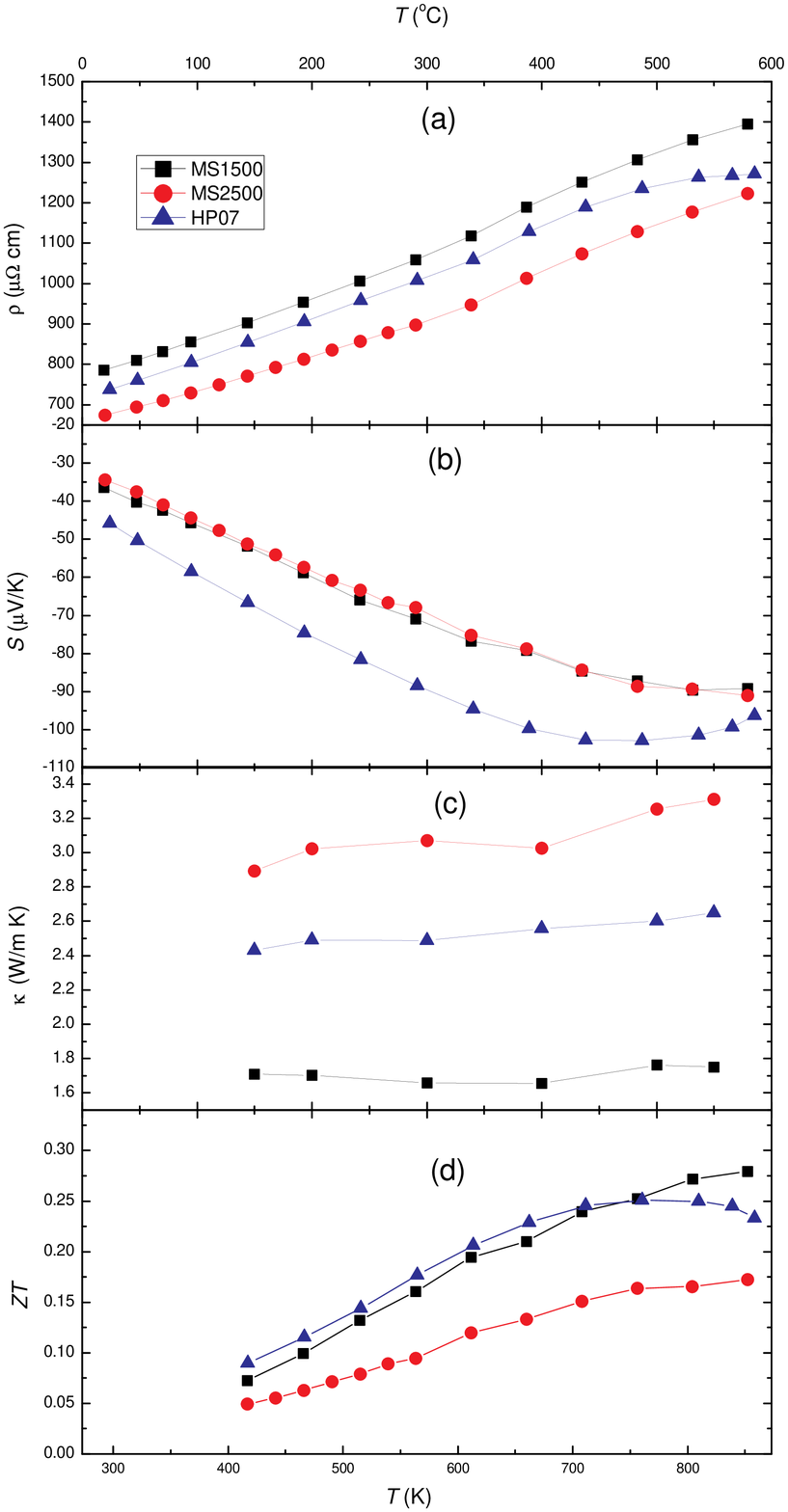}%
  \caption{\label{Fig:fig7} (Color online) Temperature dependent electrical
resistivity, $\rho(T)$, in (a), Seebeck coefficient, $S(T)$, in (b), thermal
conductivity, $\kappa(T)$, in (c) and figure of merit, $ZT(T)$, in (d) of
Ba$_8$Cu$_6$Si$_{40}$ prepared via different preparation routes (see text).}
\end{figure}

In order to further optimize $ZT$ of the clathrates in the Ba--Cu--Si system, we
melt-spun an as cast sample of Ba$_8$Cu$_6$Si$_{40}$ with two different Cu wheel
speeds, 1500 and 2500 r/min. The resulting ribbons were hand milled rather than
ball milled to retain the microstructure of the ribbons to the largest possible
extent, and then hot pressed. These samples are denoted by MS1500 and MS2500.
Figure \ref{Fig:fig7} shows $\rho(T)$, $S(T)$, $\kappa(T)$ and $ZT(T)$ of MS1500
and MS2500, together with the corresponding data for the conventionally prepared
sample HP07 of the same nominal composition. While the sample HP07 reaches the
largest absolute $S$ values in the measured temperature range, $|S(T)|$ of the
melt-spun samples is still increasing the temperature. The sample MS2500 has the
lowest $\rho$ in the measured temperature but also here a maximum in $\rho(T)$
appears to be reached at lower temperatures for HP07 than for the melt-spun
samples. MS1500 shows the lowest $\kappa(T)$. Above $480^\circ$C, MS1500 has the
highest $ZT$ value.

As seen from Fig.\,\ref{Fig:fig7}(b), $S(T)$ is very similar for MS1500 and
MS2500. This indicates that also the charge carrier concentrations are
comparable: the values $1.88 \times 10^{27}$~m$^{-3}$ for MS2500 and $1.86
\times 10^{27}$~m$^{-3}$ for MS1500 are derived from linear fits to $S(T)$ using
Eqn.\,\ref{eq:eq1} as done above. Hall effect measurements also yield similar
$n_H$ within the above used one-band model: $2.18 \times 10^{27}$~m$^{-3}$ for
MS2500 and $2.11 \times 10^{27}$~m$^{-3}$ for MS1500. The more pronounced differences between MS2500 and
MS1500 in $\rho(T)$ and $\kappa$ must therefore be attributed to the different
charge carrier mobilities $\mu_H(T) = R_H(T)/\rho(T)$, which are 2.4 and
3.5~cm$^2$/Vs at 300~K for MS1500 and MS2500, respectively. We attribute that
to a smaller average grain size in MS1500 than in
MS2500.

%
%
%
%
\subsection {DFT results}\label{theo}
\subsubsection{Structure and energetics}
   \begin{figure}
    \begin{center}
    \includegraphics[width=0.4\textwidth]{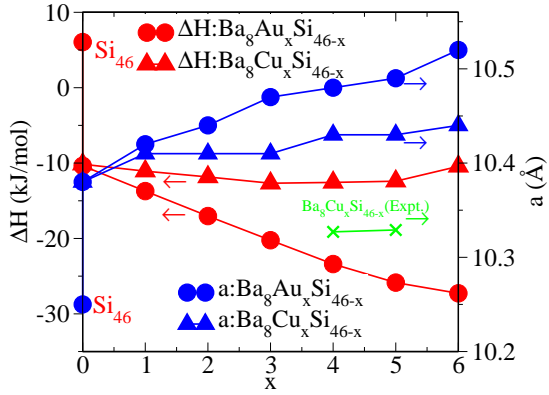}
    \end{center}
    \caption{DFT derived formation energies $\Delta H$ and lattice parameters of 
             Ba$_8$Au$_x$Si$_{46-x}$ (balls) and Ba$_8$Cu$_x$Si$_{46-x}$ (triangles)
             as a function of doping.
            Also shown are the results for the unfilled cage trcuture of Si$_{46}$.}
    \label{BaCuSiallat}
    \end{figure}
The lattice parameters for the cubic type I  clathrate structure are derived from fully relaxed VASP
calculations, from which also the enthalpies of formation $\Delta H$ in terms of differences of total
energies at zero pressure are derived. 
We compare our calculations on Ba$_8$Cu$_x$Si$_{46-x}$ with a recent study on the type I clathrate Ba$_8$Au$_x$Si$_{46-x}$.
Figure \,\ref{BaCuSiallat} shows formation enthalpies and lattice parameters for 
four types of systems , namely Si$_{46}$, 
Ba$_8$Si$_{46}$, Ba$_8$Cu$_x$Si$_{46-x}$  and Ba$_8$Au$_x$Si$_{46-x}$ with $x=1$ to 6.
From the measured X-ray data it was
concluded that the Cu atoms are randomly distributed on Si $6d$ sites due to the preparation
at elevated temperatures. 
However, for the DFT calculations the doping atoms have to be placed
on specific sites within the unit cell which, in general, causes changes of lattice symmetry
and cell shape.
Therefore, to be consistent with experiment a cubic unit cell was enforced for all calculations. 
More details are given in Refs\,\onlinecite{zeiringer_phase_2011}. 
Comparing to the two experimental data points for the lattice parameter in Fig. \ref{BaCuSiallat}, one
finds that the DFT derived values are larger by about 1\%, but show the same trend of
increasing with increasing $x$. 
The (rather small) difference between first principles
theory and experiment is attributed to the approximation for the exchange correlation functional. 
For the Au-doped compounds
the enthalpy of formation decreases (i.e. bonding is enhanced) with increasing
Au content, which is very
similar to the results for Ba$_8$Ag$_x$Ge$_{46-x}$ \cite{zeiringer_phase_2011}. 
The trend for the Cu-doped compounds is quite different, because
$\Delta H$ remains rather constant with increasing $x$. A more careful examination
reveals that
$\Delta H$ slightly increases for $x>5$
implying that a doping level $x$ exceeding 5 is unfavorable. 
This is consistent with the experimental fact
that no compound can be stabilized for such large dopings.

\subsubsection{Electronic properties}
   \begin{figure}
    \begin{center}
    \includegraphics[width=0.4\textwidth]{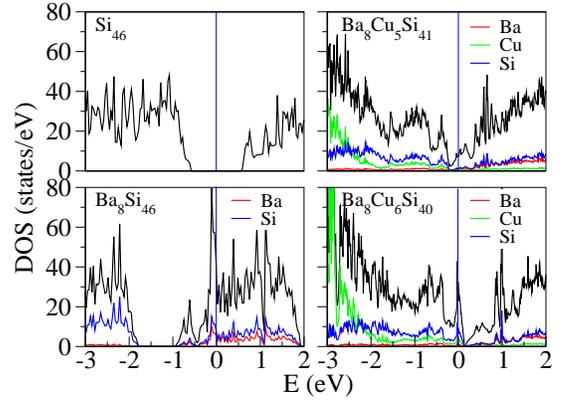}
    \end{center}
    \caption{Density of states (DOS) of Ba$_8$Cu$_x$Si$_{46-x}$ ($x$ = 0, 5 and 6). 
  Fermi energy at $E=0$ eV.}
    \label{Ba8CuxSiyPDOS}
    \end{figure}
The electronic densities of states for Ba$_8$Cu$_x$Si$_{46-x}$ are shown in Fig.\,\ref{Ba8CuxSiyPDOS}. 
Significant changes induced by doping Cu are observed. 
On filling  Ba into the voids of the Si$_{46}$ cage  the gap is maintained 
but shrinks in size due to hybridization between Ba and Si states. 
The gap decreases significantly by adding  Cu, 
which implies hybridization between dopants and the Si framework. 
For Ba$_8$Cu$_5$Si$_{41}$ there are two peaks in the DOS around the Fermi energy, 
which is located at a satellite of the higher energy peak. 
Doping of Cu up to 6 atoms lowers the higher energy peak in Ba$_8$Cu$_6$Si$_{40}$  in magnitude and 
moves the Fermi energy to the center of the peak  at  lower energy. 
The small  difference in the shape of the DOS peaks between these two compounds 
may be due to their  different crystal symmetries. 
However,  there is also a significant  similarity between Ba$_8$Cu$_5$Si$_{41}$ and Ba$_8$Cu$_6$Si$_{40}$ in that 
the peaks around the Fermi energy remain almost unchanged and also the pseudogap 
occurs for both of them. 
The similarity of the DOS features justifies  the rigid band approximation
when varying the atomic  composition in terms of changing the number of 
valence electrons by shifting the Fermi energy. For the validity of the rigid
band model the change of the number of valence electrons must be sufficiently small.

The location of the Fermi level plays an important role in understanding transport properties 
as discussed later. 
In our recent work within the rigid band approximation \cite{zeiringer_phase_2011}, based on the semiconducting parent compounds of  Ge$_{46}$
a simple electron counting rule was proposed: $x_{gap} = \frac{16}{4-n}$ 
($n$ is the valency of the dopant) is replaced the Fermi energy in the gap for Ba$_8$M$_x$Ge$_{46-x}$
. 
This counting rule can also be applied to Ba$_8$M$_x$Si$_{46-x}$ because of the
similarity of Si and Ge.
For M = Cu (i.e. the valency $n$ equals 1)  $x_{gap}$ = 16/3 is derived, 
which corresponds very well to experimental concentrations at which the Seebeck coefficient
undergoes a drastic change in size.
\subsubsection{Transport properties}
For small variations of the dopant concentration (i.e. variation $\Delta n$ of the number of valence electrons) 
the rigid band approximation was employed according to
  \begin{equation}
   N=\int_{-\infty}^{\infty}g(E)f(T,\mu)dE
  \end{equation}
for calculating the chemical potential $\mu$ (i.e. the Fermi energy) for $N=N_0+\Delta n$ electrons.
The number $N_0$ represents the number of valence electrons of the chosen reference compound with 
its DOS, $g(E)$. 
In the Ba--Cu--Si clathrates, Ba, Cu and Si contribute 2, 1 and 4 valence electrons to the system, 
respectively. 
For comparing the calculations with measurements on the most interesting experimental sample 
with composition Ba$_8$Cu$_{4.9}$Si$_{41.1}$,  $\Delta n=0.3$
if Ba$_8$Cu$_5$Si$_{41}$ is chosen as the reference and 
$\Delta n$ = 3.3 if Ba$_8$Cu$_6$Si$_{40}$ is the reference.

   \begin{figure}
    \begin{center}
    \includegraphics[width=0.4\textwidth]{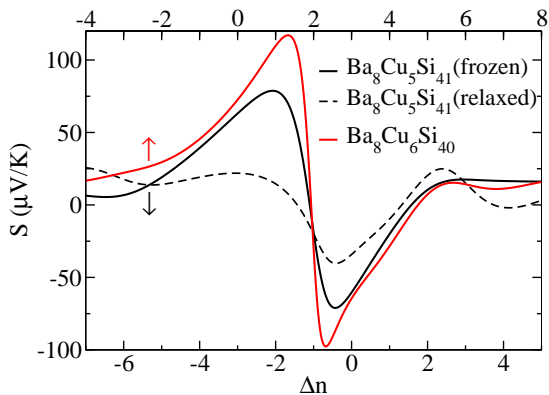}
    \end{center}
    \caption{Calculated Seebeck coefficient at 300 K for Ba$_8$Cu$_5$Si$_{41}$ and
Ba$_8$Cu$_6$Si$_{40}$ as a function of doping $\Delta n$. A negative value corresponds to hole doping.}
    \label{BaCuSi_S_N}
    \end{figure}
Figure\,\ref{BaCuSi_S_N} depicts the calculated Seebeck coefficients at
300 K vs. chemical doping for relaxed- and frozen-Ba$_8$Cu$_5$Si$_{41}$ and Ba$_8$Cu$_6$Si$_{40}$, 
which change sign as the Fermi level crosses the gap.
There is a significant difference in the amplitudes of the Seebeck coefficient between the relaxed and 
frozen structures of Ba$_8$Cu$_5$Si$_{41}$ revealing the influence of
structural relaxation.
The Seebeck coefficient of frozen-Ba$_8$Cu$_5$Si$_{41}$ resembles quite well that of Ba$_8$Cu$_6$Si$_{40}$,
indicating that the rigid band approximation should be useful when varying $x$
between 5 and 6.
Both curves agree very well with each other except for the absolute values nearby the gap. 
However, the Seebeck coefficients are significantly different for the fully relaxed structures.
This is different from the comparison of Seebeck coefficients of Ba$_8$Ag$_5$Si$_{41}$ and
Ba$_8$Ag$_6$Si$_{40}$ with their fully relaxed structures,
because these practically coincide.
\cite{zeiringer_phase_2011} 

   \begin{figure}
    \begin{center}
    \includegraphics[width=0.4\textwidth]{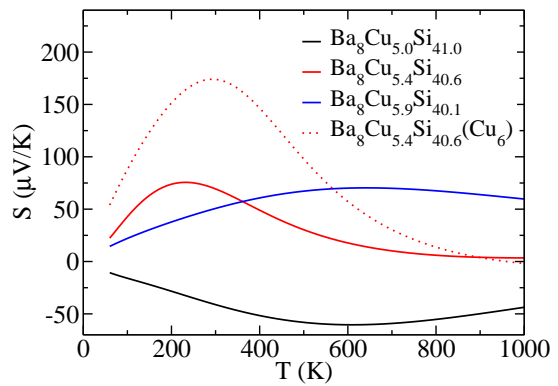}
    \end{center}
    \caption{DFT derived Seebeck coefficients for Ba$_8$Cu$_x$Si$_{46-x}$ ($x=5.0-5.9$)
    as a function of temperature as calculated within the rigid band approximation
     taking frozen-Ba$_8$Cu$_5$Si$_{41}$ as the reference. 
     For $Ba_8Cu_{5.4}Si_{40.6}$ the Seebeck coefficient with
Ba$_8$Cu$_6$Si$_{40}$ as the reference is also shown for comparison(dotted line).}
    \label{Ba8Cu5Si41seeb1}
    \end{figure}

   \begin{figure}
    \begin{center}
    \includegraphics[width=0.4\textwidth]{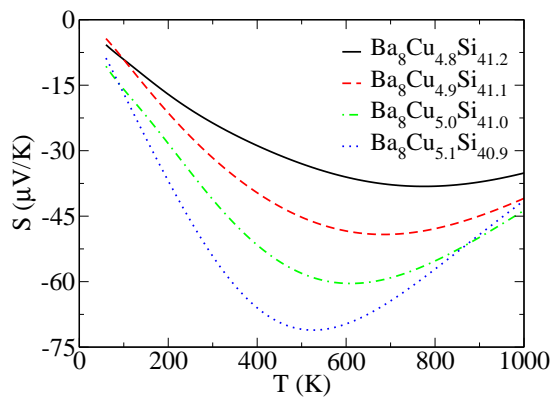}
    \end{center}
    \caption{DFT derived Seebeck coefficients for Ba$_8$Cu$_x$Si$_{46-x}$ ($x=4.8-5.1$)
    as a function of temperature as calculated within the rigid band approximation
     for which frozen-Ba$_8$Cu$_5$Si$_{41}$ is the reference.}
    \label{Ba8Cu5Si41seeb2}
    \end{figure}

Figures \ref{Ba8Cu5Si41seeb1} and \ref{Ba8Cu5Si41seeb2} 
depict the calculated temperature-dependent Seebeck coefficients $S(T)$
for Ba$_8$Cu$_x$Si$_{46-x}$, which are obviously very sensitive to doping.
For comparison, $S$ for Ba$_8$Cu$_{5.4}$Si$_{40.6}$ was also
calculated by taking Ba$_8$Cu$_6$Si$_{40}$ as the reference.
Although  $S(T)$ for Ba$_8$Cu$_{5.4}$Si$_{40.6}$ as derived from the different
references are qualitatively similar, there are
significant differences concerning  the magnitude and the position of the
maximum.  This suggests that care has to be taken for selecting the reference compound.
Apparently, $S(T)$ at low temperatures changes sign when the Cu concentration
crosses the critical value of $x_{gap}$ = 5.33.
The change of sign for Ba$_8$Cu$_{x}$Si$_{46-x}$ 
can be understood in terms of the Mott's formula,
which is a simplification valid at very low temperatures. There,
the sign and magnitude of the Seebeck coefficient 
are determined by the DOS and its derivative with respect to energy at the Fermi energy $E_F$, 
$-\frac{dg(E)}{dE}\frac{1}{g(E)}|_{E_F}$,
which results  in an opposite sign for the  Seebeck coefficient and the slope of the DOS at $E_F$.
According to Mott a small $g(E_F)$ together with a large $\frac{dg(E)}{dE}|_{E_F}$ 
gives rise to a large Seebeck coefficient, and therefore
the Fermi energy should be as close as possible to the gap. This consideration
is, however, only valid at sufficiently low temperatures.
As temperature increases, the Fermi energy moves towards or  away from the gap, 
thus changing  $S(T)$. 

From Fig.\,\ref{Ba8CuxSiyPDOS} one derives a positive derivative for Ba$_8$Cu$_{5}$Si$_{41}$ 
thus producing a negative Seebeck coefficient.
For compositions close to
Ba$_8$Cu$_5$Si$_{41}$, for instance Ba$_8$Cu$_{4.9}$Si$_{41.1}$,
the valence electron number is 1.3 electron larger than
the critical value $x_{gap}$ = 5.33, which means
that the Fermi energy is above the gap, higher than
the Fermi energy of Ba$_8$Cu$_5$Si$_{41}$. 
Further decreasing the Cu concentration, i.e., increasing the number of valence
electrons, shifts the Fermi energy away from the gap,
which significantly reduces the absolute value of $S(T)$
although the change in the number of valence electrons is
rather small. 

Summarizing, the DFT results for $S(T)$ resemble the experimental
trend rather well.  It should be noted that rather small uncertainties in the
stoichiometry, which are unavoidable in the preparation of the samples,
result in significant variations of the number of valence
electrons, which concomitantly leads to substantial variations
of the Seebeck coefficients, as demonstrated by
Fig.\,\ref{Ba8Cu5Si41seeb2}.

   \begin{figure}
     \begin{center}
     \includegraphics[width=0.4\textwidth]{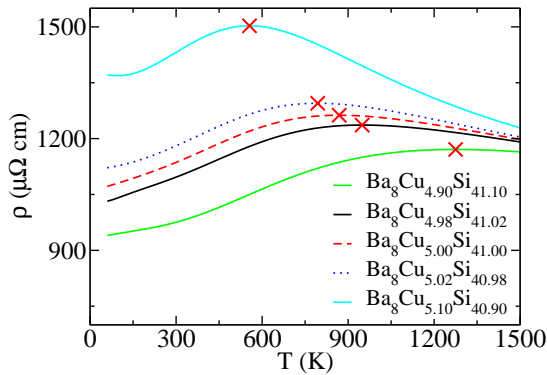}
     \end{center}
     \caption{DFT derived electrical resistivities of Ba$_8$Cu$_x$Si$_{46-x}$ 
     within the rigid band approximation with frozen-Ba$_8$Cu$_5$Si$_{41}$ as the reference.
     Crosses denote the positions of the maxima.}
     \label{Ba8Cu5Si41Rfrom6}
    \end{figure}
In Fig.\,\ref{Ba8Cu5Si41Rfrom6} the resistivity $\rho(T)$ is shown 
for Ba$_8$Cu$_x$Si$_{46-x}$ with $x=4.90-5.10$, 
as derived for frozen-Ba$_8$Cu$_5$Si$_{41}$ as the reference.
The relaxation time $\tau$ = 2.32 $\times$ $10^{-14}$ s was chosen 
to fit the calculated $\rho(T)$ of Ba$_8$Cu$_{5.0}$Si$_{41.0}$ to 
the experimental maximum of $\rho(T)$ of Ba$_8$Cu$_{4.9}$Si$_{41.1}$.
Inspecting Fig.\,\ref{Ba8Cu5Si41Rfrom6} one realizes that the calculations yield a trend similar
to the experimental data, because
the characteristic peak shifts toward higher temperatures with  decreasing $x$.
The small difference in $x$ between theory and experiment, when requiring good
agreement,  is attributed to the experimental uncertainties of sample
preparation.
The maximum  in the electrical resistivity is closely related to the position of the Fermi energy. 
In recent work on Ba$_8$Ag$_x$Ge$_{46-x}$, it was found \cite{zeiringer_phase_2011} that the peak
is due to a particular position of the Fermi level in combination with  
the temperature dependence of the energy derivative of the Fermi function, $\frac{df(E)}{dE}$
. 
More specifically,  if the Fermi energy is above but very close to the gap, 
the states below the gap will contribute to the electronic transport only
for sufficiently large temperatures when $\frac{df(E)}{dE}$ is sufficiently
broad.
As the Fermi energy moves closer or away from the gap, 
concomitantly lower or higher temperatures are required for thermally exiting states below the gap. 
From the  discussion above, 
one derives  that the Fermi energies for all the compounds are above the gap and 
are approaching it  as $x$ increases  from 4.9 to 5.1. 

From the discussion above we conclude,
that the thermoelectric properties and 
the conductivity of Ba$_8$Cu$_x$Si$_{46-x}$ alloys can be reasonably described 
and understood by means of a DFT approach.
\section{Conclusion}

In summary, thermoelectric properties of the ternary clathrate
Ba$_8$Cu$_x$Si$_{46-x}$ with the nominal Cu contents $x = 3.6, 3.8, 4.2, 4.4,
4.6, 5, 6, 7$ have been investigated. All thermoelectric quantities vary
systematically with the actual Cu content in the clathrate phase. The highest
figure of merit $ZT = 0.23$ is achieved for the sample with the nominal
composition $x = 6$ at about $450^\circ$C. Almost the same $ZT$ is retained over
a relatively broad temperature range from 350 to $550^\circ$C. 
For the optimal Cu content, samples were fabricated also by
melt spinning. One of these samples reaches an even larger $ZT$ value of 0.28
at $600^{\circ}$C.
Extensive density functional theory calculations have been performed. 
For deriving enthalpies of formation the respective total energies for Ba$_8$Si$_{46}$,
Ba$_8$Cu$_x$Si$_{46-x}$ and  Ba$_8$Au$_x$Si$_{46-x}$ were computed. Theroelectric properties were
 treated within Boltzmann's theory
and the properties in terms of Seebeck coefficients and resistivities
were calculated by exploiting the electronic structure of
Ba$_8$Cu$_6$Si$_{40}$ and Ba$_8$Cu$_5$Si$_{41}$. Overall, the agreement
between the density functional data and experiment is reasonable, but some
discrepancies remain to be solved.

\begin{acknowledgments}   
The authors would like to thank M.\ Wass for assistance in SEM/EDX measurements,
A. Grytsiv and M. Falmbigl for help with hot pressing, and A. Prokofiev for fruitful discussions. This work was
supported by the FFG project THECLA (815648), and partially by the FWF projects
P19458-N16 and TRP 176-N22. 
M.X.C. gratefully acknowledges the support by the FWF within the Science College
W4 ``Computational Materials Science''.
\end {acknowledgments}

\providecommand{\noopsort}[1]{}\providecommand{\singleletter}[1]{#1}%

\end{document}